\documentclass[sigconf]{acmart}

\usepackage{lineno}
\usepackage{pgfplots}
\usepackage{natbib}
\usepackage{amsmath,amsfonts}
\usepackage{algorithmic}
\usepackage{graphicx}
\usepackage{textcomp}
\usepackage{xcolor}
\usepackage{booktabs}
\usepackage{lipsum}
\usepackage{makecell}
\usepackage{multirow}
\usepackage{subcaption}
\usepackage[symbol]{footmisc}
\usepackage{longtable}
\usepackage{url}

\AtBeginDocument{%
  }

\setcopyright{acmlicensed}
\copyrightyear{2025}
\acmYear{2025}
\acmDOI{XXXXXXX.XXXXXXX}
\acmConference[WEBSCI '25]{ACM Web Science Conference}{May 20-24, 2025}{New Brunswick, New Jersey, USA}
\acmISBN{978-1-4503-XXXX-X/18/06}

\begin{document}

\title{Can Generative Agent-Based Modeling Replicate the Friendship Paradox in Social Media Simulations?}


\author{Gian Marco Orlando}
\affiliation{%
  \institution{University of Naples Federico II}
  \country{}
  }
\email{gianmarco.orlando@unina.it}

\author{Valerio La Gatta}
\affiliation{%
  \institution{Northwestern University}
  \country{}
}
\email{valerio.lagatta@northwestern.edu}

\author{Diego Russo}
\affiliation{%
 \institution{University of Bergamo}
 \country{}
 }
 \email{diego.russo@unibg.it}

\author{Vincenzo Moscato}
\affiliation{%
  \institution{University of Naples Federico II}
  \country{}
  }
  \email{vincenzo.moscato@unina.it}

\renewcommand{\shortauthors}{Orlando et al.}

\begin{abstract}
  Generative Agent-Based Modeling (GABM) is an emerging simulation paradigm that combines the reasoning abilities of Large Language Models with traditional Agent-Based Modeling to replicate complex social behaviors, including interactions on social media. While prior work has focused on localized phenomena such as opinion formation and information spread, its potential to capture global network dynamics remains underexplored. This paper addresses this gap by analyzing GABM-based social media simulations through the lens of the Friendship Paradox (FP), a counterintuitive phenomenon where individuals, on average, have fewer friends than their friends. We propose a GABM framework for social media simulations, featuring \emph{generative agents} that emulate real users with distinct personalities and interests. Using Twitter datasets on the US 2020 Election and the QAnon conspiracy, we show that the FP emerges naturally in GABM simulations. Consistent with real-world observations, the simulations unveil a hierarchical structure, where agents preferentially connect with others displaying higher activity or influence. Additionally, we find that infrequent connections primarily drive the FP, reflecting patterns in real networks. These findings validate GABM as a robust tool for modeling global social media phenomena and highlight its potential for advancing social science by enabling nuanced analysis of user behavior.
\end{abstract}




\maketitle

\section{Introduction}

Large Language Models (LLMs) have enabled the development of \emph{generative agents}, autonomous computational entities capable of reasoning, adapting, and interacting in ways resembling human behavior. Generative Agent-Based Modeling (GABM) leverages these capabilities and integrates LLM with traditional Agent-Based Modeling (ABM) to enhance simulations by enabling more sophisticated and adaptive agent behavior \cite{Park_GenAgents}. Unlike traditional ABM, which relies on probabilistic equations and oversimplifies the complexities of human decision-making \cite{pastor2024large}, GABM leverages LLMs to enable \emph{generative agents} whose behaviors arise naturally from their interactions within the simulated environment, unconstrained by predefined rules. This paradigm shift has demonstrated its potential in simulating complex systems involving human decision-making and social interactions \cite{Park_GenAgents, Epidemic_Modeling}.

Recent studies in the social sciences have explored how GABM can be applied to model human behaviors in order to investigate a wide range of micro- and meso-level phenomena \cite{STM_Decay_Formula, News_Feed_GenAgents}, primarily focusing on information propagation \cite{GenAgents_S3} and influence dynamics \cite{Social_Norms_GenAgents}. Despite these advancements, the exploration of network formation processes and global network-level phenomena within the GABM framework remains underexplored. To our knowledge, only \cite{ASONAM2024} demonstrates that \emph{generative agents} naturally form homogeneous ideological clusters, akin to real-world phenomenon of echo chambers. This paper addresses this notable research gap by focusing on the Friendship Paradox (FP), a phenomenon where individuals, on average, have fewer friends than their friends do \cite{feld1991}, due to skewed degree distributions typical of social networks. The Generalized Friendship Paradox (GFP) extends this concept to other metrics, including (i) \emph{Activity Paradox}, where individuals are, on average, less active than their friends \cite{Friendship_Paradox_Redux}; (ii) \emph{Virality Paradox}, where individuals are, on average, less influential than their friends \cite{Friendship_Paradox_Redux}; and (iii) \emph{Susceptibility Paradox}, where individuals' friends are, on average, more susceptible to influence than the individuals themselves \cite{Susceptibility_Paradox}. The FP and its variations have practical implications for content virality, opinion formation, and information asymmetry \cite{Qualities_and_Inequalities_GFP}, as well as psychological ramifications, significantly affecting individuals' perceptions \cite{Psychology_FP1}. Given these implications, exploring the reproduction of the FP and GFP within GABM offers valuable insights into its ability to model complex network-level phenomena, highlighting its potential as a powerful tool for studying online human behavior.

We propose a GABM-based social media simulation framework featuring \emph{generative agents} powered by LLaMa 3 \cite{dubey2024llama}, integrating realistic personalities, dynamic connections, memory mechanisms, and standard social media actions, alongside a recommender system to guide interactions. Our study investigates two research questions:

\begin{itemize}
    \item[\textbf{RQ1:}] \emph{Do the FP and GFP  manifest in GABM-based social media simulations?}
    \item[\textbf{RQ2:}] \emph{Which social connections most contribute to experiencing the FP and GFP?}
\end{itemize}

Given that FP and GFP have been extensively studied in real-world social networks, our primary contribution is to explore whether these network-level phenomena naturally emerge in GABM-based social media simulations. Using Twitter datasets on the US 2020 Election \cite{US_Dataset} and the QAnon conspiracy \cite{QAnon_Dataset}, we confirm the consistent emergence of the FP and its variations across simulations, with agents preferentially following those with superior attributes like higher activity or influence. Notably, in line with real social media \cite{Qualities_and_Inequalities_GFP, Online_FP2}, the FP is predominantly driven by less frequent interactions, rather than more frequent connections with closer followers. These findings not only validate the applicability of GABM in replicating complex social phenomena but also provide a new lens for understanding the mechanisms underlying the FP and GFP. This study offers valuable implications for social science research, particularly in exploring the hierarchical and asymmetric nature of social relationships in digital environments.

\section{Methodology} \label{sec: Materials_and_Methods}

\begin{figure*}
    \centerline{\includegraphics[width=0.85\linewidth]{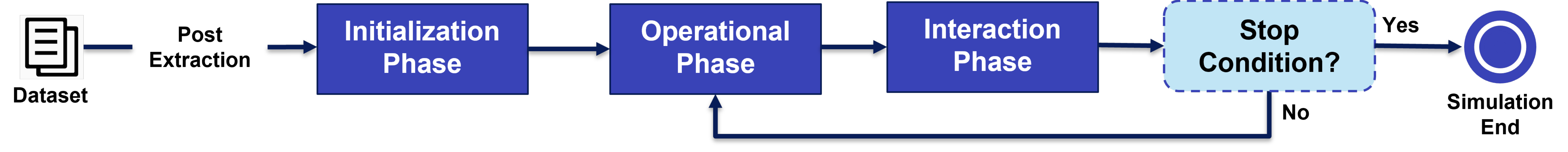}}
    \caption{The simulation workflow begins with the Initialization Phase, setting up agents with distinct personalities. In the Operational Phase, agents autonomously decide their actions, such as posting or interacting with content. The Interaction Phase updates the simulation environment. The process repeats iteratively until the stop condition is met.}
    \label{Framework_Workflow}
\end{figure*}

Our framework simulates social media dynamics through generative agents, advanced LLM-powered entities designed to simulate realistic user behavior on social media. Each agent is modeled with a profile, a memory unit, and a reasoning module that enables decision-making. Over multiple iterations, agents engage in typical social media actions such as following users, creating posts, or re-sharing content. The simulation workflow, as illustrated in Figure \ref{Framework_Workflow}, consists of three phases which collectively simulate the dynamics of the social network. We detail generative agents and each phase of the simulation workflow in the following sections.

\subsection{Generative Agents}

\begin{figure*}
    \centering
    \includegraphics[width=0.85\linewidth]{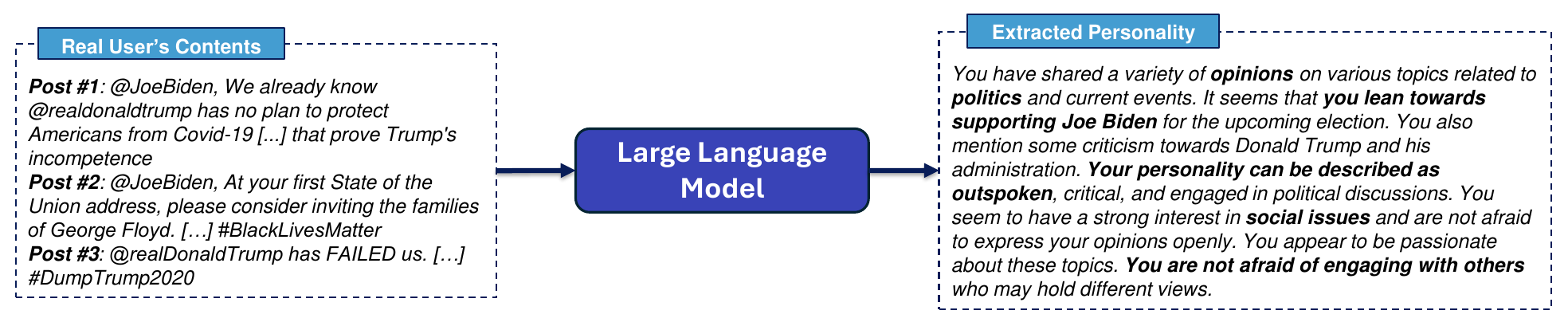}
    \caption{\textbf{Example of personality inference.} LLMs analyze user's generated content to extract personality traits.}
    \label{fig:agent_modules}
\end{figure*}

\paragraph*{\textbf{Agent Profile}}

The agent’s profile replicates a social media user’s identity, including inferred personality traits and dynamic social connections. Personality traits are generated from real user posts using a prompt-based strategy with LLMs, identifying attributes such as ideological alignment and engagement style (e.g., being outspoken, critical, or supportive of specific political figures), as illustrated in Figure \ref{fig:agent_modules}. In addition to personality profiling, each agent maintains a dynamic list of social connections in a social media graph \( \mathcal{G} = (\mathcal{V}, \mathcal{E}) \), where \( \mathcal{V} = \{ a_1, a_2, \dots, a_n \} \) represents the set of agents, and \( \mathcal{E} = \{ (a_i, a_j) \mid a_i, a_j \in \mathcal{V} \} \) denotes the directed edges capturing follower-followee relationships. While personality traits remain static, connections evolve as agents decide to follow others during the simulation.

\paragraph*{\textbf{Memory Unit}}

The agent’s activity history is managed through a memory unit with two components: Short-Term Memory (STM) and Long-Term Memory (LTM), designed to balance recent and significant interactions. The STM stores recently published content along with engagement metadata (e.g., the number of re-shares, likes, dislikes and comments). To prevent the STM from becoming overloaded or dominated by outdated content, a decay mechanism removes less relevant items over time. The probability of removing an item is defined as \( p(M_j^{a_i}) = 1 - \frac{p_j + r_j}{2} \), where \( p_j \) represents engagement popularity and \( r_j \) measures recency.

The LTM preserves high-impact content identified from STM through periodic evaluations. Items with engagement scores exceeding a threshold \( \tau \) (set to 0.5 in experiments) are transferred to LTM, ensuring the retention of only the most influential contents.

Both memory components initialize empty, allowing agents’ memories to evolve dynamically through the simulation, providing an unbiased foundation for analyzing phenomena like the Friendship Paradox.

\paragraph*{\textbf{Reasoning Module}}

The Reasoning Module drives the agent’s decision-making by leveraging LLMs to emulate user behavior using a prompt-based approach. The module outputs a \emph{Choice-Reason-Content} triplet, where \emph{Choice} specifies the selected action, \emph{Reason} details the contextual rationale, and \emph{Content} defines the associated material, such as a new post or an interaction with another agent’s post. If no action is taken, the \emph{Content} field remains empty.

\subsection{Simulation Workflow}

\paragraph*{\textbf{Initialization Phase}}

The \emph{Initialization Phase} configures generative agents to emulate real users by assigning distinct personality profiles, identifying key interests and traits as outlined in the Agent Profile section. This approach avoids the extensive parameter calibration of traditional ABM \cite{lim2022modeling} and improves on GABM methods relying solely on demographic attributes to characterize agents \cite{Epidemic_Modeling, STM_Decay_Formula}.

\paragraph*{\textbf{Operational Phase}}

During each simulation iteration, agents are immersed in an environment resembling a social media platform, with the \emph{Operational Phase} executing the core dynamics of the agent-based model. This phase focuses on constructing the input prompt for the agent's \emph{Reasoning Module}, enabling decision-making for subsequent actions by incorporating information a real user would likely consider before taking action. This prompt incorporates three key elements: \textit{(i)} feedback on the agent’s previously published content, \textit{(ii)} a personalized feed of recommended posts, and \textit{(iii)} a list of available actions the agent can perform.

Specifically, the \emph{feedback section} provides the agent with insights into how its previously published content was received by others. This feedback is retrieved from the agent's memory unit and includes raw posts along with their engagement metrics (e.g., the number of re-shares, likes, dislikes, and comments received). The \emph{recommended posts section} simulates a social media feed, presenting posts from other agents for potential interaction, such as liking, disliking, sharing, or commenting, guided by a recommendation algorithm that mimics existing social media recommender systems. The \emph{actions' list section} outlines available actions: (i) \emph{publishing a new content}, (ii) \emph{re-share}, (iii) \emph{like}, (iv) \emph{dislike} or (v) \emph{comment} posts published by other agents, (vi) \emph{refraining from interaction}.

The Operational Phase ends when the agent's Reasoning Module outputs the \emph{Choice-Reason-Content} triplet for each agent.

\paragraph*{\textbf{Interaction Phase}}

The Interaction Phase enables agent interactions, simulating a dynamic social network by updating three key elements: \textit{(i)} the \emph{follower-followee network}, \textit{(ii)} feedback on agents’ published content, and \textit{(iii)} recommended content for potential interactions.

First, the \emph{follower-followee network}, represented as a global social media graph \( \mathcal{G} = (\mathcal{V}, \mathcal{E}) \), evolves dynamically through a "who to follow" mechanism. In this process, agents are prompted to evaluate potential new connections. To manage the limited context window of LLMs, candidate suggestions are restricted to a subset of agents identified via semantic similarity. Specifically, for a target agent \( a_i \), the system calculates the average cosine similarity between the semantic embeddings of \( a_i \)'s posts and those of other agents, selecting the top-10 agents with the highest similarity scores. Semantic embeddings are computed using BERT \cite{devlin2019bert}.

Second, the Interaction Phase provides agents with \emph{feedback} on their previously published content, including metrics like re-shares, likes, dislikes, and comments. This information is stored in the agent's memory unit, forming the basis for informed decision-making in subsequent actions. By reflecting content engagement and virality, this mechanism realistically simulates the influence of community reactions on social media users.

Finally, \textit{recommended content} is generated using a Retrieval-Augmented Generation (RAG) strategy \cite{ASONAM2024}, which dynamically suggests posts aligned with agents’ preferences. The retriever queries an external vector database, updated continuously with all content published by agents in the simulation, ensuring relevant and up-to-date recommendations. By decoupling retrieval from generation, the RAG approach maintains a stable prompt size, ensuring scalability and consistent performance throughout the simulation.

\paragraph*{\textbf{Stop Condition}}

The iterative process between \emph{Operational Phase} and \emph{Interaction Phase} continues until a predefined stop condition is met. To mitigate the decline in content diversity observed in LLM-empowered agents during repeated interactions \cite{padmakumar2023does}, we define a \emph{saturation point} determined by monitoring post similarity across iterations, marking the point where agents cease generating novel content. The simulation halts when $L$ original posts — corresponding to the number of agents — exhibit a cosine similarity greater than 0.99 with prior content. This threshold ensures the system stops only when redundancy dominates, signaling a lack of substantive new contributions.


\section{Experiments} \label{sec: Experiments}

\paragraph{\textbf{Datasets}}

We assess our framework using two publicly available Twitter datasets: \textit{(i)} the US 2020 Election dataset \cite{US_Dataset}, which captures polarized discussions between liberals and conservatives during the final months of the 2020 presidential campaign and its aftermath, reflecting a broad and polarized discourse involving more than 1 million users; \textit{(ii)} the QAnon dataset \cite{QAnon_Dataset}, which focuses on interactions within the ideologically cohesive QAnon community leading up to the January 6th Capitol riot \cite{amarasingam2020qanon}, highlighting interactions within a smaller, tightly-knit network of approximately 175,000 users. This diversity enables the examination of interactions within individual communities as well as across polarized groups, demonstrating the generalizability of our framework across varied online environments.


\paragraph{\textbf{Experimental setup}}

The experiments were conducted on a computing system featuring an 11th Gen Intel Core i7-11800H processor, 16 GB of RAM, and an NVIDIA GeForce RTX 3060 Laptop GPU. Agents were instantiated by annotating the political or ideological affiliations of 100 users per dataset, categorized as Republican or Democratic for the US election dataset and QAnon promoters or non-promoters for the QAnon dataset. Simulations concluded upon generating 100 repeated original contents, corresponding to the number of agents, indicating a saturation point where all agents had, on average, contributed to content repetition. The framework, implemented using PyAutogen \cite{AutoGen}, incorporates ChromaDB\footnote{https://www.trychroma.com/} as the vector database for the agents' memory unit and the RAG technique, and employs the open-source Llama 3 8B model\footnote{https://ai.meta.com/blog/meta-llama-3/}.

\paragraph{\textbf{Measuring the Friendship Paradox}}

To examine the Friendship Paradox and its generalizations, we analyze multiple metrics reflecting social connectivity and influence dynamics. Specifically, \emph{In-Degree} (number of followers) and \emph{Out-Degree} (number of followees) quantify the agent's social connections and enable the verification of the Friendship Paradox as a property of the network in its classical form: \emph{"on average, your friends have more friends than you do"} \cite{feld1991}. Metrics such as the \emph{Number of Tweets} (NT), representing the number of posts generated by the agent (including both original posts and re-shares), and \emph{Number of Original Tweets} (NOT), reflecting authored posts, measure activity and enable analysis of the \emph{Activity Paradox}. Then, \emph{Total Times Retweeted} (TTR), which captures the total retweets received by the agent, and \emph{Retweets per Tweet} (RPT), the ratio of re-shares to original posts, quantify the agent's influence and address the \emph{Virality Paradox} \cite{Friendship_Paradox_Redux}. Lastly, the \emph{Influence-driven Adoption Rate} (IAR), defined as the ratio of content shared by the agent to the total content exposure, examines the \emph{Susceptibility Paradox}, where an agent’s friends exhibit greater susceptibility to influence than the agent itself \cite{Susceptibility_Paradox}. We measured the Pearson correlation coefficients among the above-mentioned attributes. The correlations are generally low in magnitude, confirming that these metrics capture distinct aspects of nodal characteristics.

The analysis of the Friendship Paradox and its generalizations can be operationalized using the concept of \emph{neighbor superiority} \cite{Neighbor_Superiority}. In a social media graph $\mathcal{G} = (\mathcal{V}, \mathcal{E})$, where $\mathcal{V} = \{ a_1, a_2, \cdots, a_n \}$ is the set of agents and $\mathcal{E} = \{(a_i, a_j) | a_i, a_j \in \mathcal{V} \} $ is the set of edges representing follower-followee relationships, let $a_i^p$ denote the $p$-th attribute of agent $a_i$, and $\eta(a_i)$ (resp. $\gamma(a_i)$) represent its followees (resp. followers). For a fixed attribute $p$, agent $a_i$ experiences neighbor superiority if $a_i^p < \mathbb{E}[\eta(a_i^p)]$ (resp. $a_i^p < \mathbb{E}[\gamma(a_i^p)]$), meaning its followees (resp. followers) have higher average $p$ values. To mitigate outlier effects, the comparison can be also compared to the median of the neighbors’ values, offering a more robust measure of neighbor superiority \cite{Qualities_and_Inequalities_GFP}.

\begin{table}[t]
    \tiny
    \centering
    \caption{Percentage of agents experiencing mean neighbor superiority. The results are averaged across three simulation runs, standard deviation is in parenthesis. Bold indicates cases where the proportion of agents experiencing mean followee superiority exceeds that of agents experiencing mean follower superiority.
    }
    \label{tab:results_of_agent_metrics}
    \resizebox{\columnwidth}{!}{%
    \begin{tabular}{lcc|cc}
        & \multicolumn{2}{c|}{\textbf{US Election Dataset}} 
        & \multicolumn{2}{c}{\textbf{QAnon Dataset}}\\ \cmidrule{2-5}
        & \begin{tabular}[c]{@{}c@{}}\textbf{Mean} \\ \textbf{Follower} \\ {(\%)}\end{tabular} 
        & \begin{tabular}[c]{@{}c@{}}\textbf{Mean} \\ \textbf{Followee} \\ {(\%)}\end{tabular}
        & \begin{tabular}[c]{@{}c@{}}\textbf{Mean} \\ \textbf{Follower} \\ {(\%)}\end{tabular} 
        & \begin{tabular}[c]{@{}c@{}}\textbf{Mean} \\ \textbf{Followee} \\ {(\%)}\end{tabular}
         \\ \midrule
        In-Degree & \makecell{41.33 \tiny{(± 6.79)}} & \makecell{\textbf{86.33} \tiny{(± 2.49)}} & \makecell{19.33 \tiny{(± 6.12)}} & \makecell{\textbf{82.33} \tiny{(± 2.62)}} \\[2ex]
        Out-Degree & \makecell{26.66 \tiny{(± 5.43)}} & \makecell{\textbf{32.00} \tiny{(± 6.97)}} & \makecell{21.33 \tiny{(± 5.43)}} & \makecell{\textbf{34.00} \tiny{(± 0.81)}} \\[2ex]
        NT & \makecell{48.33 \tiny{(± 2.35)}} & \makecell{35.33 \tiny{(± 4.49)}} & \makecell{32.00 \tiny{(± 1.41)}} & \makecell{\textbf{52.66} \tiny{(± 2.86)}} \\[2ex]
        NOT & \makecell{48.66 \tiny{(± 2.86)}} & \makecell{44.00 \tiny{(± 2.16)}} & \makecell{30.00 \tiny{(± 4.54)}} & \makecell{\textbf{60.33} \tiny{(± 4.11)}} \\[2ex]
        TTR & \makecell{48.00 \tiny{(± 12.19)}} & \makecell{\textbf{78.33} \tiny{(± 4.11)}} & \makecell{31.66 \tiny{(± 5.24)}} & \makecell{\textbf{81.66} \tiny{(± 2.05)}} \\[2ex]
        RPT & \makecell{50.00 \tiny{(± 12.19)}} & \makecell{\textbf{81.66} \tiny{(± 4.98)}} & \makecell{29.66 \tiny{(± 5.43)}} & \makecell{\textbf{81.33} \tiny{(± 2.62)}} \\[2ex]
        IAR & \makecell{41.33 \tiny{(± 4.18)}} & \makecell{\textbf{75.66} \tiny{(± 1.88)}} & \makecell{33.00 \tiny{(± 2.16)}} & \makecell{\textbf{62.33} \tiny{(± 5.43)}} \\ 
        \bottomrule
    \end{tabular}
    }
\end{table}

\subsection{RQ1: Do the FP and GFP manifest in GABM-based social media simulations?}

We assess the prevalence of the Friendship Paradox and its generalizations by evaluating the percentage of agents experiencing \emph{neighbor superiority}. Table \ref{tab:results_of_agent_metrics} presents the mean percentages of agents experiencing neighbor superiority for each attribute across all datasets, averaged over three independent simulation runs, with standard deviations reported. For instance, in the US Election dataset, an average of 86.33\% of agents experience \emph{mean followee superiority} for the \emph{in-degree} attribute, indicating that for these agents $a_i \in \mathcal{V}$, the inequality $a_i^{\text{in-degree}} < \mathbb{E}[\eta(a_i^{\text{in-degree}})]$ holds.  In contrast, only 41.33\% of agents experience \emph{mean follower superiority} for the same attribute, i.e., $a_i^{in\text{-}degree} < \mathbb{E}[\gamma(a_i^{in-degree})]$. Across all datasets and attributes, followee superiority consistently surpasses follower superiority, a pattern observed in 12 of the 14 cases in Table \ref{tab:results_of_agent_metrics}. While the analysis also holds when considering median neighbor superiority, these results are omitted here for brevity. These findings align with empirical results from real-world social networks \cite{Qualities_and_Inequalities_GFP}, indicating agents tend to follow others with higher nodal attributes. This asymmetry suggests an implicit social hierarchy, reflecting the tendency of agents to follow those with higher popularity and activity.

\subsection{RQ2: Which social connections most contribute to experiencing the FP and GFP?}

We now investigate which connections primarily drive the Friendship Paradox and its generalizations for a generic agent.

To operationalize this concept, we restrict the sets of followers and followees based on interaction frequency. Formally, let $\eta(a_i)|_k$ represent the top-$k$ followers with whom agent $a_i$ has the highest interaction frequency, measured by re-shares, likes, dislikes and comments. Similarly, $\gamma(a_i)|_k$ denotes the top-$k$ followees with whom $a_i$ engages most frequently. We then analyze neighbor superiority across three cases: \textit{(i)} $\eta(a_i)|_{k=1}$ and $\gamma(a_i)|_{k=1}$, considering only the single most interactive follower or followee; \textit{(ii)} $\eta(a_i)|_{k=3}$ and $\gamma(a_i)|_{k=3}$, focusing on the top-3 most interactive connections; and \textit{(iii)} $\eta(a_i)$ and $\gamma(a_i)$, examining the complete sets of followers and followees. This approach evaluates the role of closest connections in driving the paradoxes.

\begin{figure}[t]
    \centering
    \textbf{Mean Follower Superiority} \par\medskip
    \begin{subfigure}{0.23\textwidth}
        \centering
        \includegraphics[width=\linewidth]{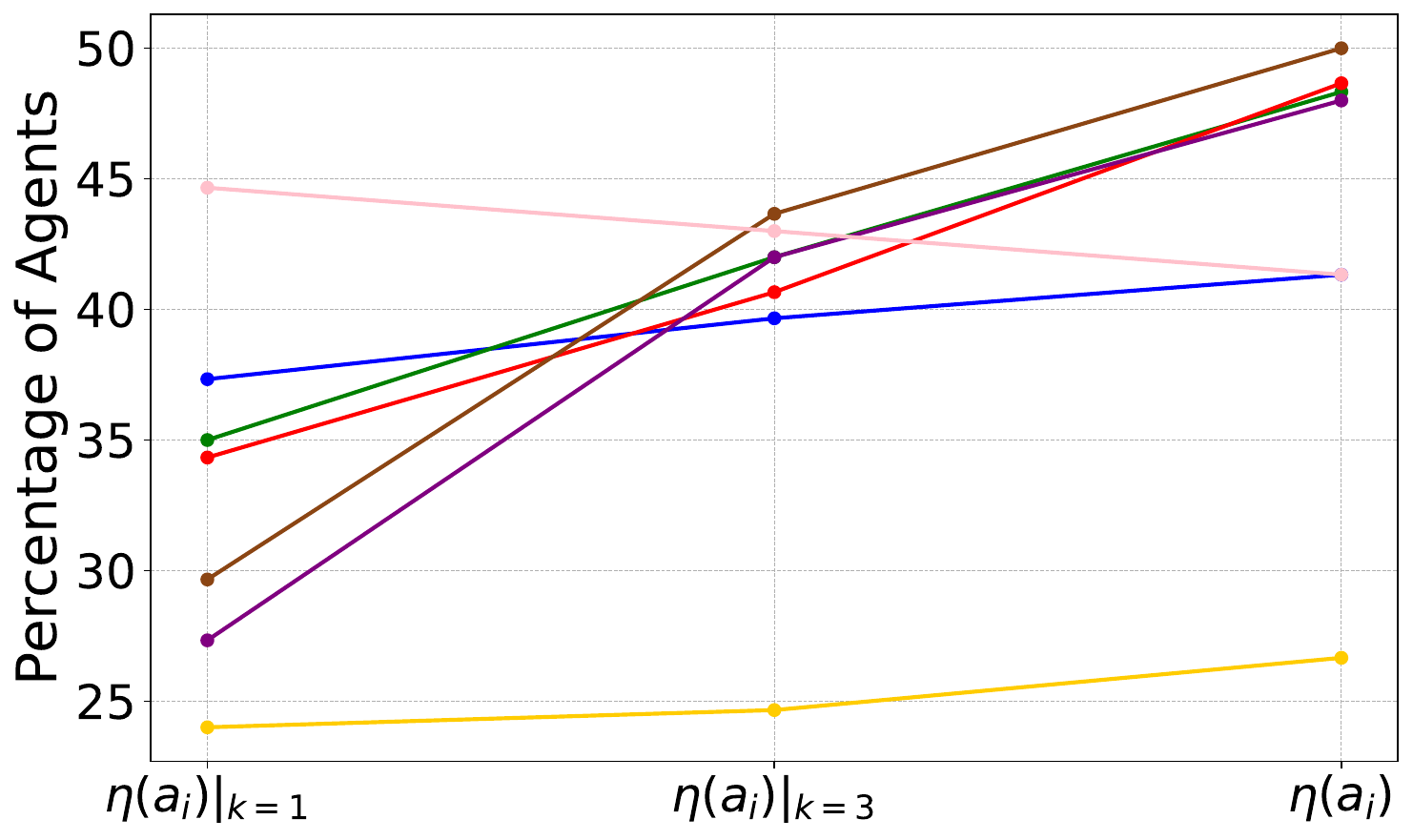}
        \caption{US Election Dataset}
        \label{USA_Election_Mean_Follower}
    \end{subfigure}
    \begin{subfigure}{0.23\textwidth}
        \centering
        \includegraphics[width=\linewidth]{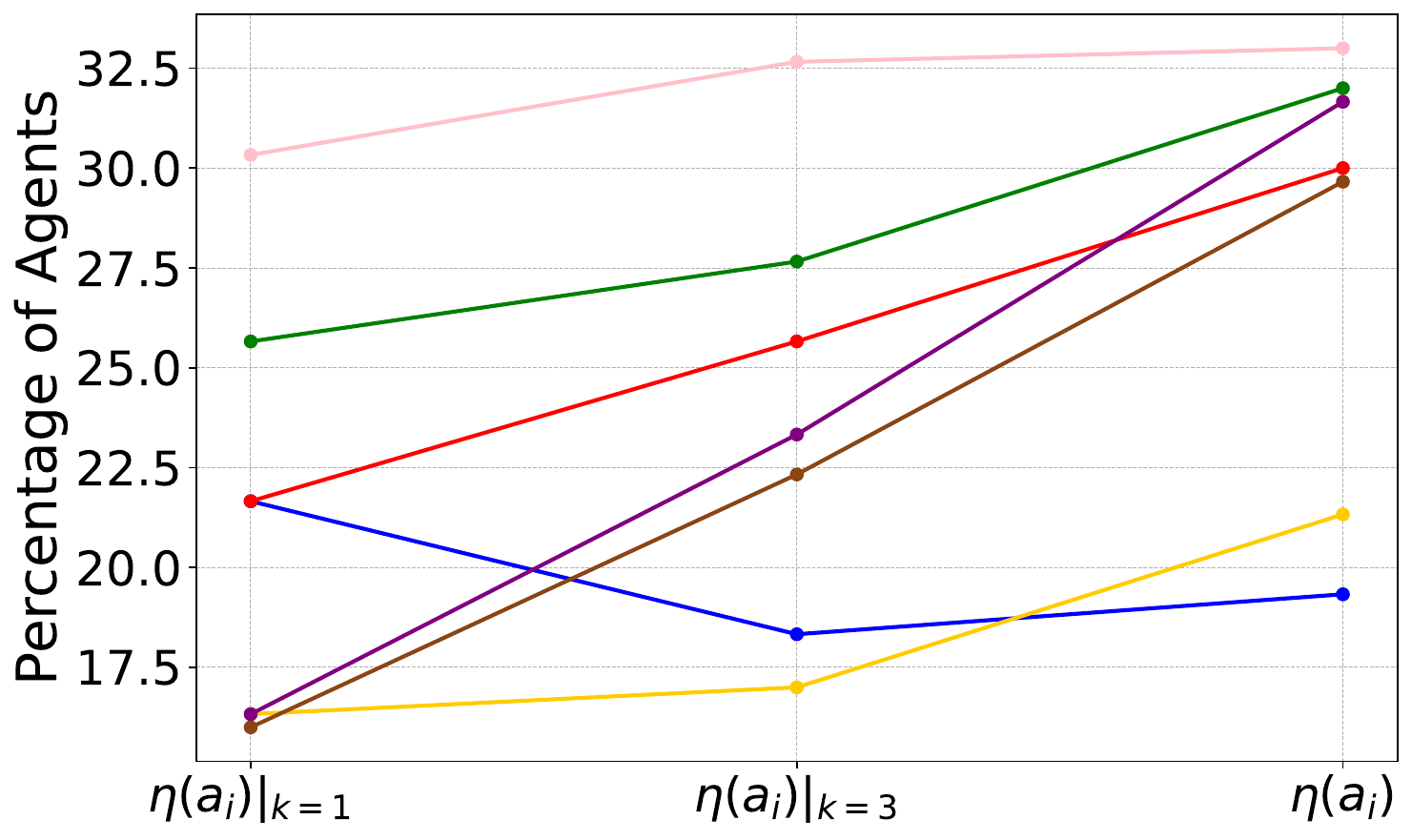}
        \subcaption{QAnon Dataset}
        \label{QAnon_Mean_Follower}
    \end{subfigure}
    \vspace{0.1cm} \par\medskip
    \textbf{Mean Followee Superiority} \par\medskip
    \begin{subfigure}{0.23\textwidth}
        \centering
        \includegraphics[width=\linewidth]{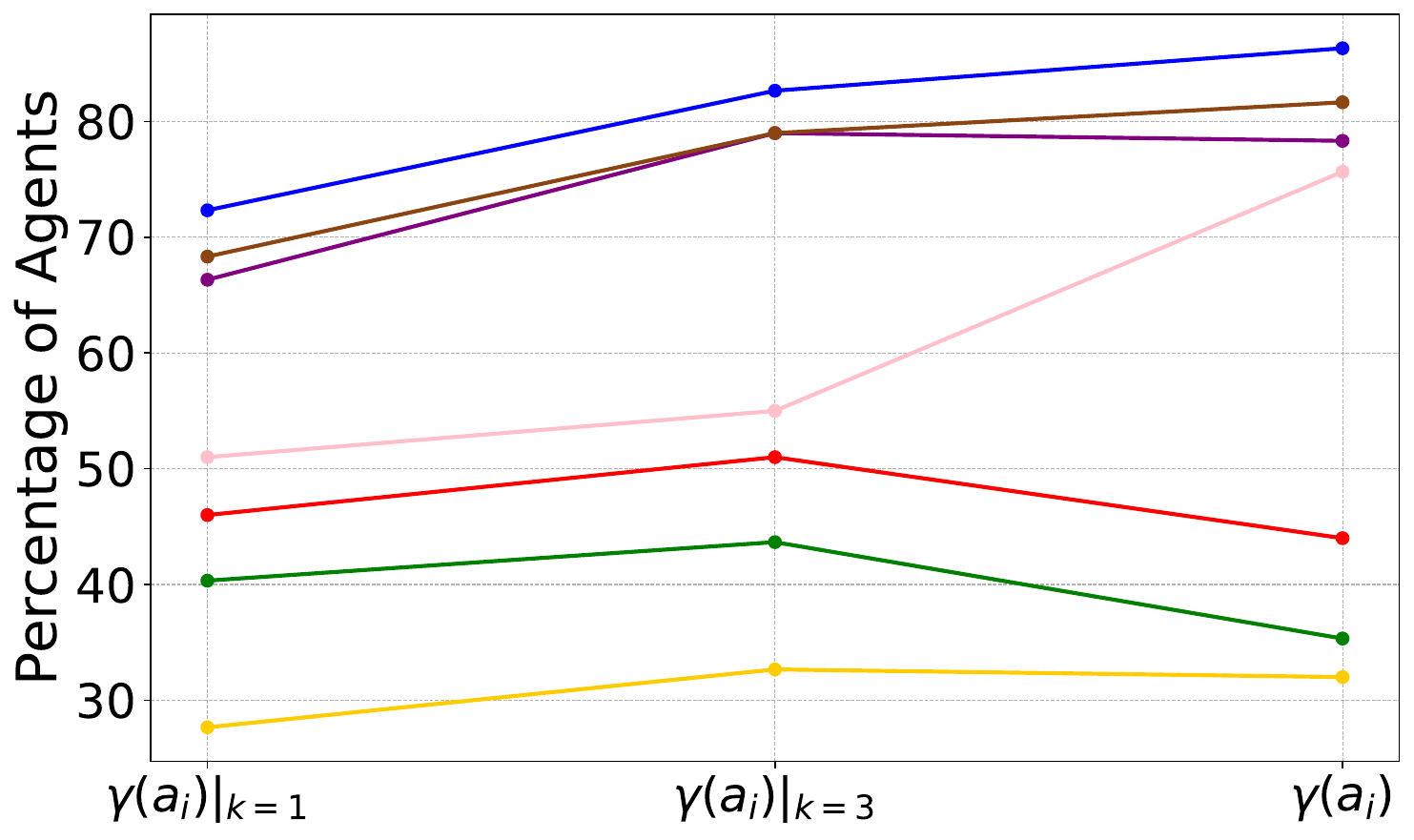}
        \caption{US Election Dataset}
        \label{USA_Election_Mean_Followee}
    \end{subfigure}
    \begin{subfigure}{0.23\textwidth}
        \centering
        \includegraphics[width=\linewidth]{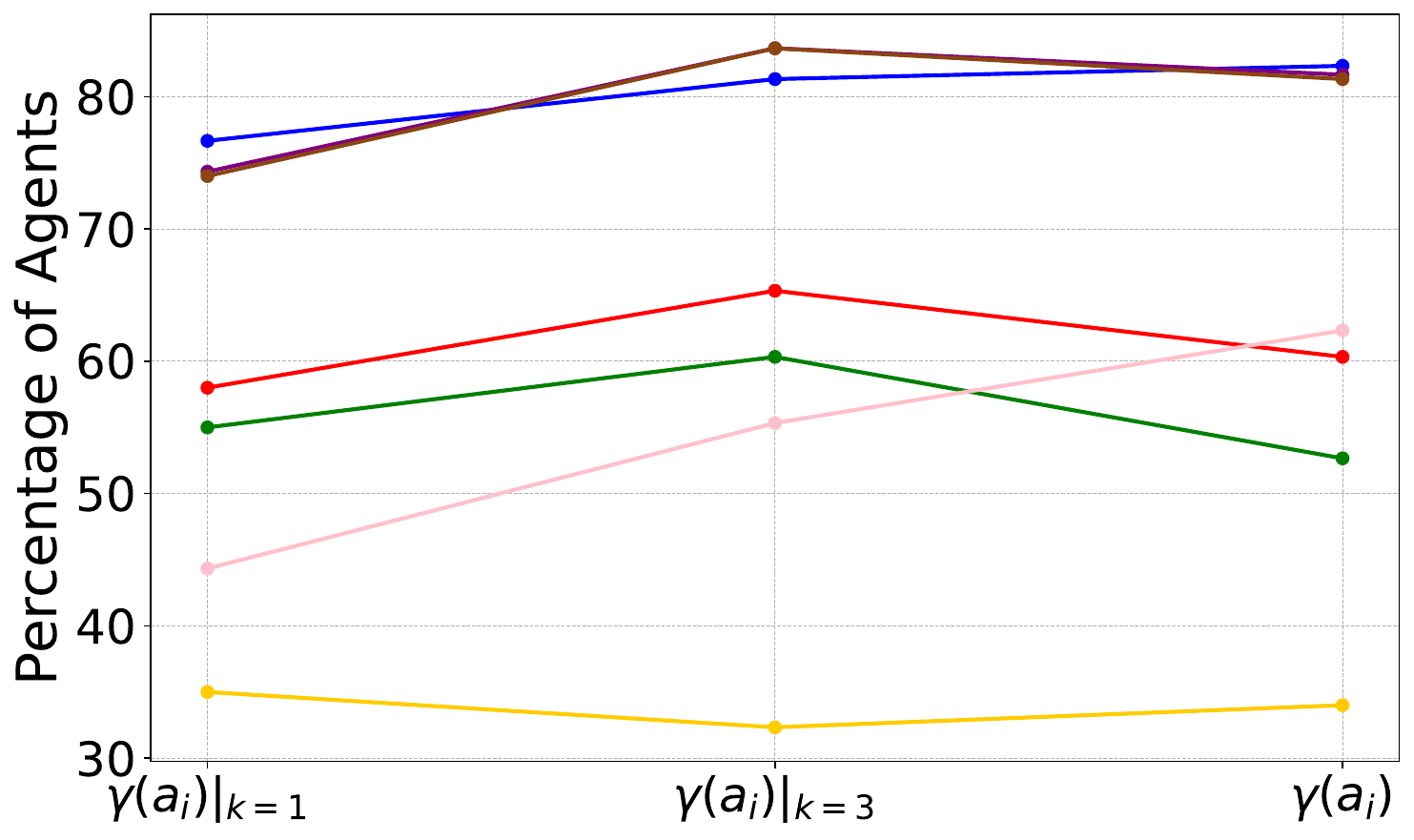}
        \caption{QAnon Dataset}
        \label{QAnon_Mean_Followee}
    \end{subfigure}
    \vspace{0.1cm} \par\medskip
    \begin{subfigure}{0.4\textwidth}
        \centering
        \includegraphics[width=\textwidth]{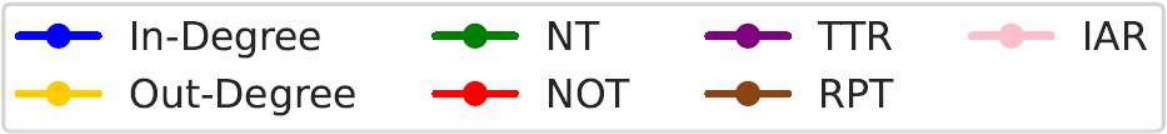}
    \end{subfigure}
    \caption{Proportion of agents experiencing mean follower superiority and mean followee superiority for each attribute across different restriction conditions.}
    \label{fig:RQ2_Results}
\end{figure}

Figure \ref{fig:RQ2_Results} shows the percentages of agents experiencing mean follower superiority (Figures \ref{USA_Election_Mean_Follower}–\ref{QAnon_Mean_Follower}) and mean followee superiority (Figures \ref{USA_Election_Mean_Followee}–\ref{QAnon_Mean_Followee}) for each attribute and restriction condition. For example, in the US Election dataset, Figure \ref{USA_Election_Mean_Followee} reveals an increasing trend in followee superiority for the in-degree attribute (blue line). Specifically, the paradox affects 72.33\% of agents under $\eta(a_i)|_{k=1}$, 82.66\% under $\eta(a_i)|_{k=3}$, and 86.33\% for the full set $\eta(a_i)$. Similarly, follower superiority (Figure \ref{USA_Election_Mean_Follower}) rises from 37.33\% under $\gamma(a_i)|_{k=1}$, increasing to 39.66\% for $\gamma(a_i)|_{k=3}$ and peaking at 41.33\% when the full set $\gamma(a_i)$ is considered. These increasing trends are consistent across most attributes for mean follower values, being observed in 17 out of 21 cases. In contrast, mean followee values exhibit this trend in approximately half of the cases (10 out of 21). These findings suggest that the Friendship Paradox and its variations are not driven by an agent’s strongest connections (i.e., those with the highest interaction frequencies). Instead, the connections that contribute to the Friendship Paradox for a given agent primarily involve agents with whom the target agent has less interactions. This aligns with real-world findings \cite{Online_FP2}, demonstrating that the paradox in GABM simulations emerges through infrequent yet impactful connections with highly influential agents.

\section{Conclusion \& Future Work} \label{sec: Conclusions}

This paper introduces a GABM framework for simulating social media dynamics. The proposed \emph{generative agents}, powered by modern LLMs, emulate real-world users by integrating distinct personalities, interests, dynamic social connections, and memory-based mechanisms. The simulation workflow enables agents to perform typical social media activities such as following, posting, re-sharing, liking, disliking, and commenting. Using this framework, we investigated the emergence of the Friendship Paradox and its generalizations, including the Activity, Virality, and Susceptibility Paradoxes, in GABM-based simulations. Experiments demonstrated that these paradoxes naturally arise in these simulations, consistent with prior findings in real-world social networks \cite{Qualities_and_Inequalities_GFP, Online_FP2}. Key findings include: \textit{(i)} agents exhibit hierarchical, preferential attachment behaviors, connecting more frequently with individuals possessing superior attributes such as higher activity or influence (\textbf{RQ1}), and \textit{(ii)} the paradoxes are primarily driven by infrequent connections rather than the most frequently interacted connections (\textbf{RQ2}). These findings validate GABM's ability to replicate complex global phenomena observed in real-world social networks.

Future work will focus on extending our study by addressing two specific areas derived from our current findings. First, we aim to investigate the temporal evolution of the FP and its variations within simulations. Analyzing these temporal trends will provide a deeper understanding of the conditions under which the paradox emerges and stabilizes. Additionally, we plan to explore which types of agents are most affected by the paradox. By pursuing these directions, we aim to further establish GABM as a novel analysis technique for social network analysis, enhancing its utility for studying online behaviors and addressing critical challenges such as misinformation and network polarization.

\begin{acks}

\end{acks}

\bibliographystyle{ACM-Reference-Format}


\end{document}